\documentclass[english,aps,manuscript]{revtex4}
\usepackage[T1]{fontenc}
\usepackage[latin9]{luainputenc}
\usepackage{amstext}
\usepackage{graphicx}

\makeatletter
\@ifundefined{textcolor}{}
{%
 \definecolor{BLACK}{gray}{0}
 \definecolor{WHITE}{gray}{1}
 \definecolor{RED}{rgb}{1,0,0}
 \definecolor{GREEN}{rgb}{0,1,0}
 \definecolor{BLUE}{rgb}{0,0,1}
 \definecolor{CYAN}{cmyk}{1,0,0,0}
 \definecolor{MAGENTA}{cmyk}{0,1,0,0}
 \definecolor{YELLOW}{cmyk}{0,0,1,0}
}

\makeatother

\usepackage{babel}
\begin{document}

\title{Reconfigurable optomechanical circulator and directional amplifier}

\author{Zhen Shen}

\affiliation{Key Laboratory of Quantum Information, Chinese Academy of Sciences,
University of Science and Technology of China, Hefei 230026, P. R.
China.}

\affiliation{Synergetic Innovation Center of Quantum Information and Quantum Physics,
University of Science and Technology of China, Hefei, Anhui 230026,
P. R. China.}

\author{Yan-Lei Zhang}

\affiliation{Key Laboratory of Quantum Information, Chinese Academy of Sciences,
University of Science and Technology of China, Hefei 230026, P. R.
China.}

\affiliation{Synergetic Innovation Center of Quantum Information and Quantum Physics,
University of Science and Technology of China, Hefei, Anhui 230026,
P. R. China.}

\author{Yuan Chen}

\affiliation{Key Laboratory of Quantum Information, Chinese Academy of Sciences,
University of Science and Technology of China, Hefei 230026, P. R.
China.}

\affiliation{Synergetic Innovation Center of Quantum Information and Quantum Physics,
University of Science and Technology of China, Hefei, Anhui 230026,
P. R. China.}

\author{Fang-Wen Sun}
\email{fwsun@ustc.edu.cn}

\affiliation{Key Laboratory of Quantum Information, Chinese Academy of Sciences,
University of Science and Technology of China, Hefei 230026, P. R.
China.}

\affiliation{Synergetic Innovation Center of Quantum Information and Quantum Physics,
University of Science and Technology of China, Hefei, Anhui 230026,
P. R. China.}

\author{Xu-Bo Zou}

\affiliation{Key Laboratory of Quantum Information, Chinese Academy of Sciences,
University of Science and Technology of China, Hefei 230026, P. R.
China.}

\affiliation{Synergetic Innovation Center of Quantum Information and Quantum Physics,
University of Science and Technology of China, Hefei, Anhui 230026,
P. R. China.}

\author{Guang-Can Guo}

\affiliation{Key Laboratory of Quantum Information, Chinese Academy of Sciences,
University of Science and Technology of China, Hefei 230026, P. R.
China.}

\affiliation{Synergetic Innovation Center of Quantum Information and Quantum Physics,
University of Science and Technology of China, Hefei, Anhui 230026,
P. R. China.}

\author{Chang-Ling Zou}
\email{clzou321@ustc.edu.cn}

\affiliation{Key Laboratory of Quantum Information, Chinese Academy of Sciences,
University of Science and Technology of China, Hefei 230026, P. R.
China.}

\affiliation{Synergetic Innovation Center of Quantum Information and Quantum Physics,
University of Science and Technology of China, Hefei, Anhui 230026,
P. R. China.}

\author{Chun-Hua Dong}
\email{chunhua@ustc.edu.cn}

\affiliation{Key Laboratory of Quantum Information, Chinese Academy of Sciences,
University of Science and Technology of China, Hefei 230026, P. R.
China.}

\affiliation{Synergetic Innovation Center of Quantum Information and Quantum Physics,
University of Science and Technology of China, Hefei, Anhui 230026,
P. R. China.}
\begin{abstract}
Non-reciprocal devices, which allow the non-reciprocal signal routing,
serve as the fundamental elements in photonic and microwave circuits
and are crucial in both classical and quantum information processing.
The radiation-pressure-induced coupling between light and mechanical
motion in traveling wave resonators has been exploited to break the
Lorentz reciprocity, realizing non-reciprocal devices without magnetic
materials. Here, we experimentally demonstrate a reconfigurable non-reciprocal
device with alternative functions of either a circulator or a directional
amplifier via the optomechanically induced coherent photon-phonon
conversion or gain. The demonstrated device exhibits considerable
flexibility and offers exciting opportunities for combining reconfigurability,
non-reciprocity and active properties in single photonic structures,
which can also be generalized to microwave as well as acoustic circuits.
\end{abstract}
\maketitle
The field of classical and quantum information processing with integrated
photonics has achieved significant progresses during past decades,
and most optical devices of the basic functionality have been realized
\cite{Fund}. Nonetheless, it is still a challenge to obtain devices
with non-reciprocal or active gain properties. Especially the non-reciprocal
devices, including the most common isolator and circulator, have attracted
great efforts for both fundamental and practical considerations \cite{Mizumoto,reservoir1,chiralnetwork,Amplifier2,Amplifier3,Amplifier4}.
Although their bulky counterparts play a vital role in daily applications
of optics, the requirement of strong external bias magnetic field
and the magnetic field shields, and also the compatibility of lossy
mageto-optics materials prevent the miniaturization \cite{BiL}.

Owing to the general principle of Lorentz reciprocity or time-reversal
symmetry in optics, nonlinear optical effects are the remain option
to come round the obstacle\cite{Fan1,Fan2,Fan3}. So far, the optical
isolation based on spatiotemporal modulations and three-wave mixing
effects have been developed \cite{Kang,brillouin1,Brillouin2,XiangGuo,wenjie,Jiangxiaoshun,rabl,OptoMech_Shen,OptoMech2,OptoMech3},
and the similar mechanism has been applied to superconducting microwave
circuits \cite{microwave1,microwave4,ElecMech1,ElecMech2,ElecMech3}.
Recently, the fiber-integrated optical circulator for single photon
has also been realized, in which nonreciprocal behavior arises from
the chiral interaction between the atom and the transversely confined
light \cite{atomcirculator}. However, the optical circulator and
the directional amplifier for large dynamic range of signal power
is still inaccessible. Here, we demonstrate the first optomechanical
circulator and directional amplifier in a two-tapered fiber coupled
silica microresonator, which is used as an add-drop filter \cite{adddrop1,Adddrop2}.
Via a simple change in control field, this add-drop filter can be
switched into circulator mode or directional amplifier mode. Our
device has several advantages beyond its bulky counterparts, includes
the reconfigurability, amplification, compactness.

\begin{figure}
\includegraphics[clip,width=0.8\columnwidth]{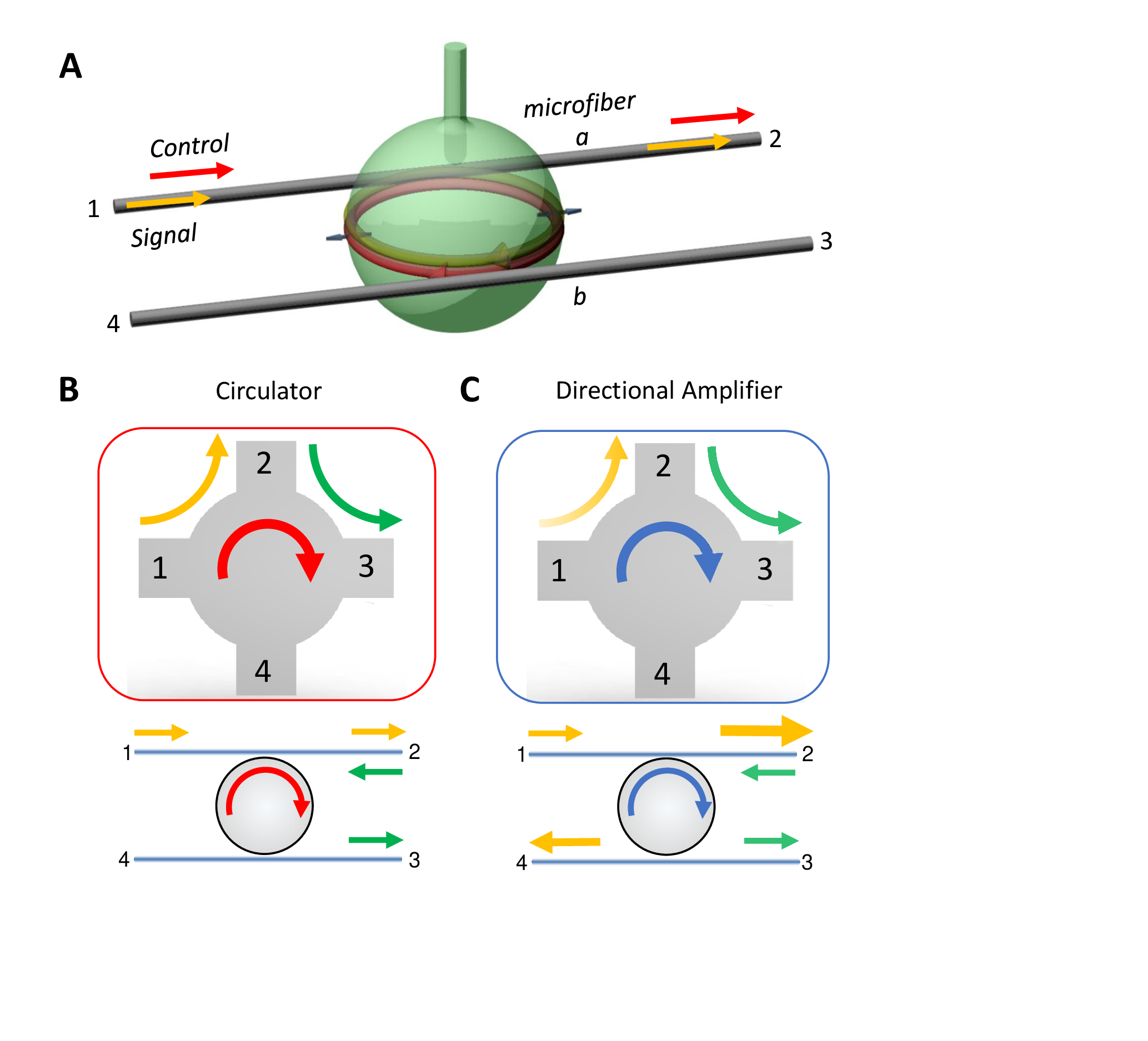}\caption{Schematic of the optomechanical circulator and directional amplifier.
(A) The device consists of an optomechanical resonator and two coupled
microfibers. A control field launched into port 1 excites the coupling
between the mechanical motion and the clockwise optical field. (B
and C) The routing direction of the signal light coincides with the
control field (i.e., clockwise direction). The port number follows
that of microfibers in (A). For the directional amplifier, the arrow
color getting deeper (or size becoming larger) indicates gain in the
corresponding direction, while the arrows remaining unchanged represent
no gain.}
\end{figure}

The optomechanical circulator and directional amplifier feature the
photonic structure as in Fig.$\,$1A, where a silica microsphere resonator
is evanescently coupled with two tapered microfibers (designated $a$
and $b$) as signal input-output channels. For a passive configuration
that without pump, it is a four-port add-drop filter device \cite{adddrop1,Adddrop2},
which can filter the signal from fiber $a$ to $b$ or vice versa
via the passive cavity resonance. Because of the traveling wave nature,
the microresonator supports pairs of degenerate clockwise (CW) and
counter-clockwise (CCW) whispering-gallery modes, and the device transmission
function is symmetric under the $1\leftrightarrow2$ and $4\leftrightarrow3$
commutation. The key to the reconfigurable non-reciprocity is the
nonlinear optomechanical coupling, represented by the Hamiltonian
\begin{equation}
H_{int}=g_{0}(c_{\mathrm{cw}}^{\dagger}c_{\mathrm{cw}}+c_{\mathrm{ccw}}^{\dagger}c_{\mathrm{ccw}})(m+m^{\dagger}),
\end{equation}
where $c_{\mathrm{cw\left(ccw\right)}}$ and $m$ denote the bosonic
operators of the CW(CCW) optical cavity mode and mechanical mode,
respectively. The radial breathing mode that changes the circumference
of the microsphere could modulate the cavity resonances, with the
$g_{0}$ is the single-photon optomechanical coupling rate.

Biased by a control cavity field that detuned from the resonance,
either the coherent conversion or the parametric coupling between
the signal photon and phonon could be enhanced \cite{Optomechreview}.
However, the bias control field can only stimulate the interaction
between phonon and signal photon that propagating along the same direction
as the bias. As a result of a directional control field, which is
chosen as CW mode in our experiment, the time-reversal symmetry is
broken and effective non-reciprocity is produced for the signal light.
In particular, the device performs the function of either a circulator
or a directional amplifier, which is determined by the frequency detuning
of the control light with respect to the cavity resonance.

When the CW optical mode is excited via a red-detuned control field,
i.e., $\omega_{c}-\omega_{o}\approx-\omega_{m}$, where $\omega_{c}$,
$\omega_{o}$, $\omega_{m}$ are the respective frequencies of the
control field, optical and mechanical modes, it gives rise to the
well-known photon-phonon coherent conversion as a beam-splitter-like
interaction $(c_{\mathrm{cw}}^{\dagger}m+c_{\mathrm{cw}}m^{\dagger})$
\cite{OMIT1,OMIT2}. For the CW signal photons sent to the cavity
through the fiber port $1(3)$ in Fig.$\,$1A, a transparent window
appears in the transmittance from port $1(3)$ to port $4(2)$ when
the signal is near resonance with the optical cavity mode. The signal
is routed by the control field due to a destructive interference between
the signal light and mechanically up-converted photons from the control
field \cite{OMIT1,OMIT2,OMIT3}. In contrast, for the two other input
ports 2 and 4, the signal light couples to the CCW optical mode and
drops to ports $3$ and $1$, respectively. Thus the add-drop functionality
is remained for these two ports since the absence of optomechanical
interaction. In general, the device functions as a four-port circulator,
in which the signal entering any port is transmitted to the adjacent
port in rotation, as shown in Fig.$\,$1B.

For a control field that is blue-detuning from the CW mode ($\omega_{c}-\omega_{o}\approx\omega_{m}$),
an effective photon-phonon pair generation process $(c_{\mathrm{cw}}^{\dagger}m^{\dagger}+c_{\mathrm{cw}}m)$
leads to signal amplification. Similar to the case of circulator,
only the signal that is launched into certain direction can couple
with mechanical mode and be amplified, as shown in Figs.$\,$1C. For
example, signal input at port $1$ leads to amplified signal output
at both port $2$ and $4$. In reverse, signal input at port $2$
will only drop to port $3$ without amplification. Therefore, such
a device can operate as an usual add-drop filter, circulator, or directional
amplifier by programming the control field.

\begin{figure}
\includegraphics[clip,width=0.8\columnwidth]{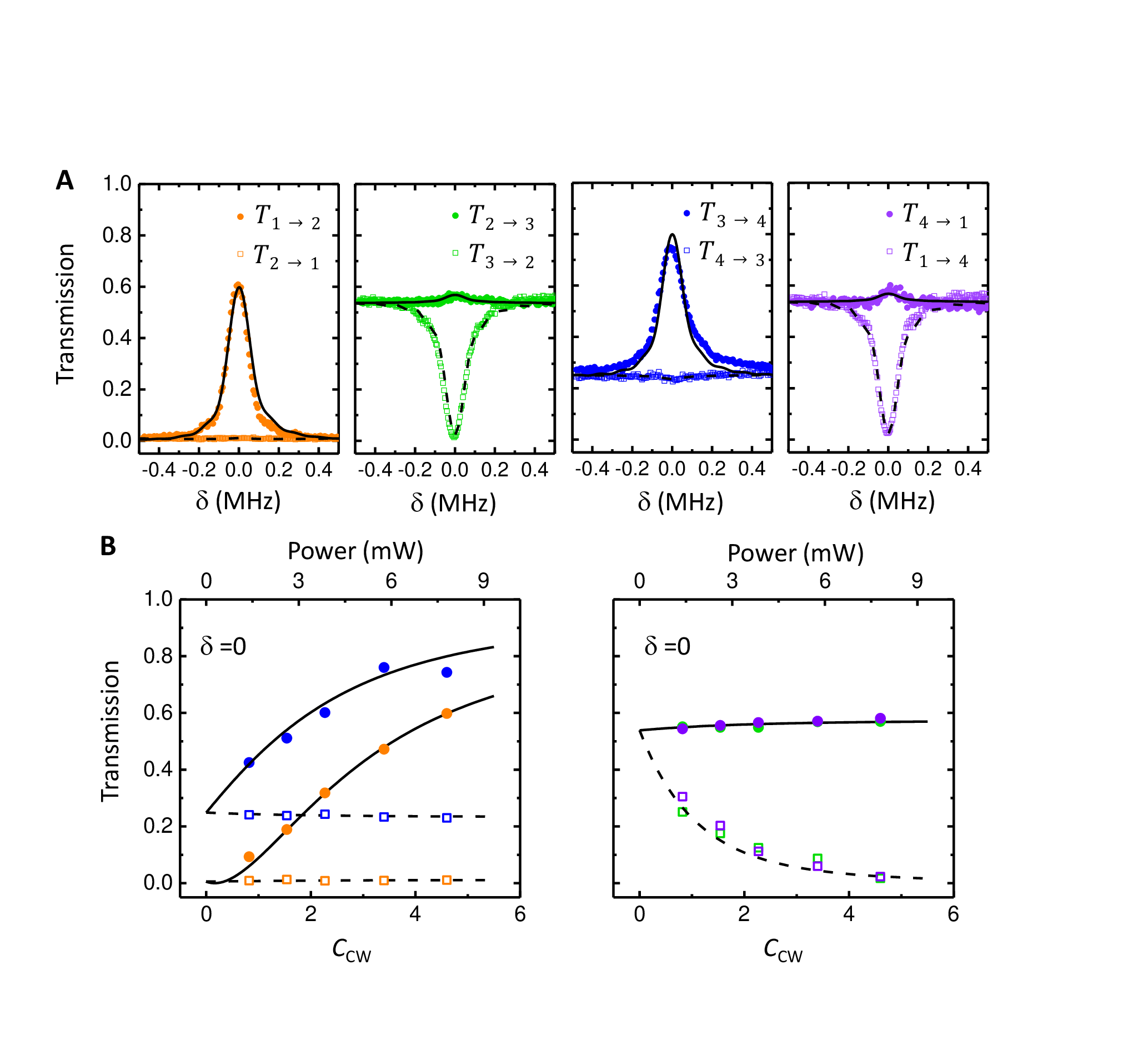}\caption{The demonstration of circulator function with red-detuned control
field. (A) The measured port-to-port transmission spectra of signal
around the cavity resonance, the solid circles for $T_{i\rightarrow i+1}$
and the open squares for $T_{i+1\rightarrow i}$. The incident control
power is 7.8 mW, corresponding to $C_{\textrm{cw}}=4.6$. (B) The
transmittance obtained at $\delta=0$ versus $C_{\textrm{cw}}$. The
lines in (A) and (B) are the theoretically expected values using the
parameters $\kappa/2\pi=16.2\:\textrm{MHz}$, $\omega_{m}/2\pi=90.47\,\textrm{MHz}$,
$\gamma/2\pi=22\:\textrm{kHz}$.}
\end{figure}

The optomechanical resonator used in this study is a silica microsphere
with a diameter of around $35\:\mu\textrm{m}$, where we choose a
high-Q-factor whispering-galley mode with intrinsic damping rate $\kappa_{0}/2\pi=3\:\textrm{MHz}$
near $780\,\mathrm{nm}$. The radial breathing mechanical mode has
a frequency of $\omega_{m}/2\pi=90.47\,\textrm{MHz}$ and a dissipation
rate of $\gamma/2\pi=22\:\textrm{kHz}$. The two microfibers are mounted
in two 3D stages and the distance between the resonator and microfibers
is fixed throughout the experiment, and the external coupling rates
of the two channels are $\kappa_{a}/2\pi=9\:\textrm{MHz}$ and $\kappa_{b}/2\pi=4.2\:\textrm{MHz}$,
respectively (see \cite{SM} for more details about the setup).

For the experimental demonstration of optomechanical circulator, we
first measure the signal transmission spectra $T_{i\rightarrow i+1}$
from $i$-th to ($i+1$)-th port and the reversal $T_{i+1\to i}$
with $i\in\left\{ 1,2,3,4\right\} $ (as shown in Fig.$\,$2A) when
the CW optical mode is excited by a red-detuned control laser. Here,
the control laser and signal light are pulsed (pulse width $\tau=10\:\mu\textrm{s}$)
to avoid the thermal instability of the microsphere \cite{OptoMech_Shen,OMIT3}.
With the detuning $\delta$ between the signal and cavity field (see
\cite{SM} for more spectra), the spectra unambiguously present asymmetric
transmittance in the forward ($i\to i+1$) and backward ($i+1\to i$)
directions around $\delta\approx0$: relatively high forward transmittance
($60\%\sim80\%$) while near-zero backward transmittance. Such performance
indicates an optical circulator (Fig.$\,$1B) with insertion loss
of around $1\sim2\,\mathrm{dB}$. The isolation for the backward $T_{4\to3}$
is slightly higher, because of the imperfection imposed by the unbalanced
external coupling rates of two channels. To achieve a better understanding
of the role of the optomechanical interactions, we measure the transmission
spectra under different intensities of the control field. The transmissions
at $\delta=0$ are summarized and plotted in Fig.$\,$2B as a function
of the cooperativity $C_{\mathrm{cw}}\equiv4g_{0}^{2}N_{d}/\kappa\gamma$,
where $N_{d}$ is the CW intracavity control photon number, and $\kappa=\kappa_{0}+\kappa_{a}+\kappa_{b}$
is the total cavity damping rate. By increasing $C_{\mathrm{cw}}$,
we observe the non-reciprocal transmittance contrast between forward
and backward directions ($T_{i\rightarrow i+1}-T_{i+1\to i}$) increases
from $0$ to around $60
$.

\begin{figure}
\includegraphics[clip,width=0.8\columnwidth]{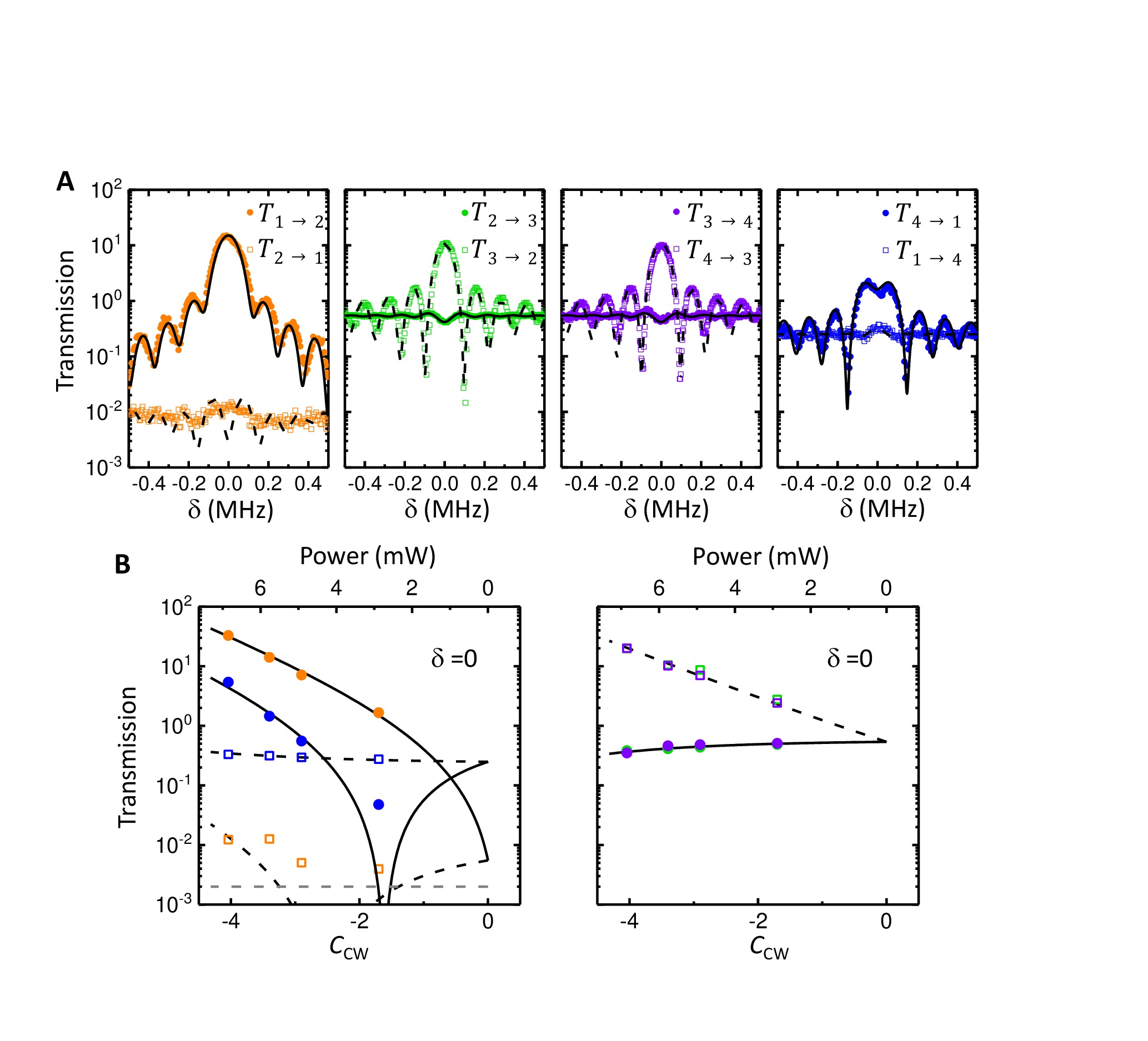}

\caption{The demonstration of directional amplifier with blue-detuned control
field. (A) The typical measured transmission spectra for the function
of directional amplifier. The solid circles represent $T_{i\rightarrow i+1}$
and the open squares represent $T_{i+1\rightarrow i}$. The incident
control power is 5.8 mW, corresponding to $C_{\mathrm{cw}}=-3.4$.
 (B) The transmittance obtained at $\delta=0$ versus $C_{\mathrm{cw}}$.
The lines in (A) and (B) are the theoretically expected values using
the parameters $\kappa/2\pi=16.2\:\textrm{MHz}$, $\omega_{m}/2\pi=90.47\,\textrm{MHz}$,
$\gamma/2\pi=22\:\textrm{kHz}$.}
\end{figure}

By tuning the frequency of control field to the upper motional sideband
of the optical mode ($\omega_{c}-\omega_{0}=\omega_{m}$) and keeping
other conditions the same as before, the same device is reconfigured
to be a directional amplifier. As shown in Fig.$\text{\,}$3A, only
the signal light launched into port 1 and port 3 (i.e., coupled to
CW mode) will be simultaneously transferred to port 2 and port 4 with
considerable gain, but not vice versa (see \cite{SM} for more spectra).
For the channel from port 2 to port 1, the lower transmittance predicted
by theory at $\delta=0$ is not measured due to the noise. Here, the
experimental results are fitted to the transient transmission spectra,
the sinc-function-like oscillations around the central peak are observed
due to the impulse response of the device for a $10\:\mu\textrm{s}$
rectangular control pulse. For the transmittance at $\delta=0$ and
$C_{\mathrm{cw}}=-4.0$, where the negative sign of cooperativity
represents the blue-detuned drive \cite{OptoMech_Shen}, the signal
field from port 1 to port 2 is amplified by 15.2 dB but in the reversed
direction it suffers 19.1 dB loss as shown in Fig.$\,$3B. Hence,
the maximum contrast ratio between forward and backward probe transmission
is approximately 34.3 dB when $C_{\mathrm{cw}}=-4.0$.

To fully characterize the performance of our reconfigurable non-reciprocal
devices, we measure the complete transmission spectra $T_{i\rightarrow j}$
between all ports (i.e., $i,$ $j\in\{1,2,3,4\}$), with $C_{\mathrm{cw}}=0$
for add-drop filter, $C_{\mathrm{cw}}>0$ for circulator and $C_{\mathrm{cw}}<0$
for directional amplifier, respectively. Figure 4 shows the experimental
results of the transmittance matrix for $C_{\mathrm{cw}}=0,\:4.6,\:-4.0$
at $\delta=0$, and also the matrix for ideal circulator as a comparison.
To quantify the device performance, we introduce the ideality metric
$I=1-\frac{1}{8}\sum_{i,j}\left|T_{i\rightarrow j}^{N}-T_{i\rightarrow j}^{I}\right|$
 , with $T_{i\rightarrow j}^{N}=T_{i\rightarrow j}/\eta_{i}$ is
the normalized experimental transmittance at $\delta=0$ for subtracting
the influence of the insertion loss (see \cite{SM} for more details),
$T_{i\rightarrow j}^{I}$ for ideal performance and $\eta_{i}=\sum_{j}T_{i\rightarrow j}$
is total output for the signal field entering port $i$. As shown
in Fig.$\,$4E, the ideality of circulator and amplifier approaches
unity with increasing $\left|C_{\mathrm{cw}}\right|$, which agrees
well with theoretical fittings \cite{SM}. The best idealities of
all three functions of the device exceed $75
$.

\begin{figure}
\includegraphics[clip,width=0.8\columnwidth]{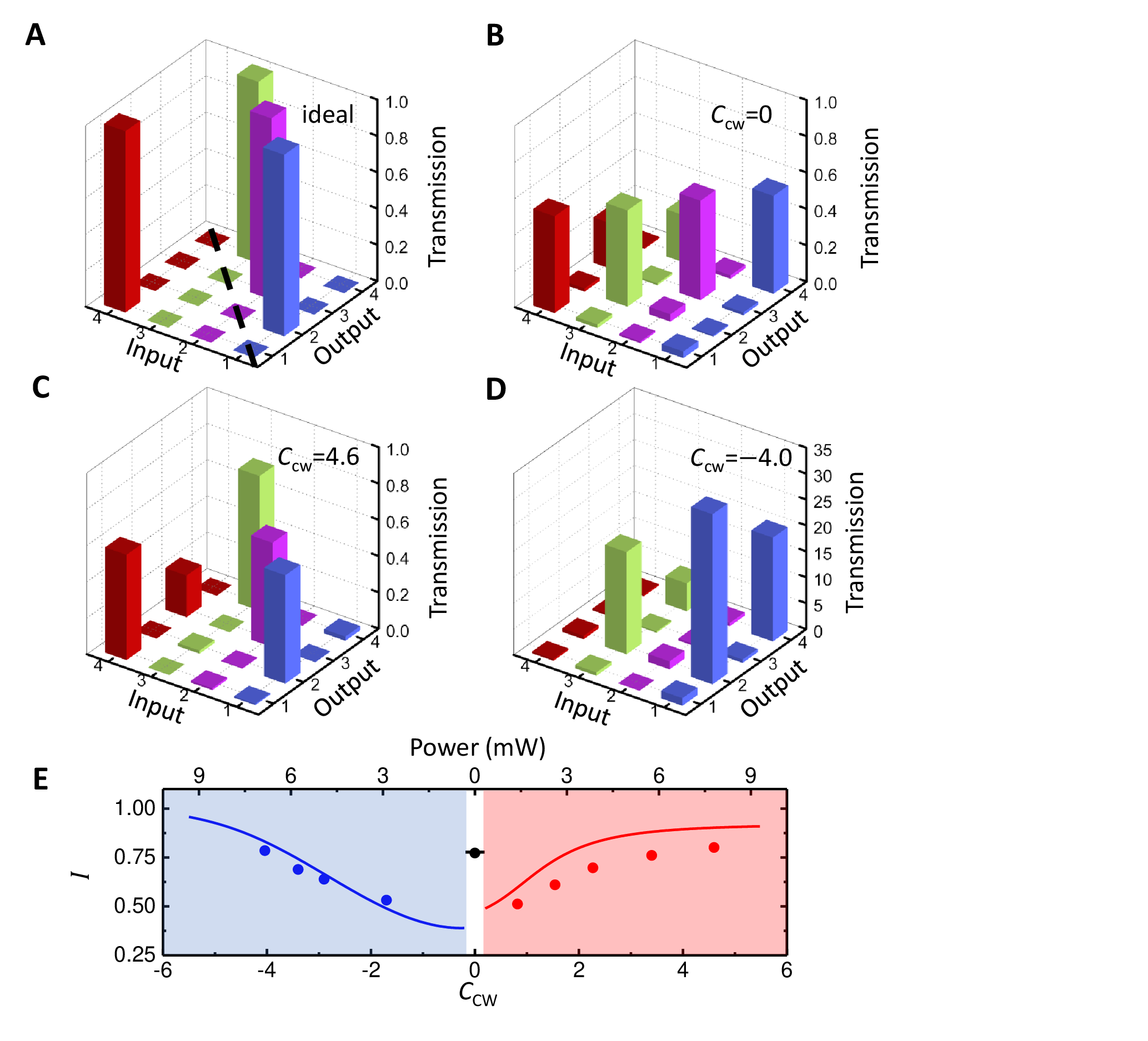}\caption{Transmission matrices. (A) The transmission matrix of an ideal circulator:
only $T_{1\rightarrow2}=T_{2\rightarrow3}=T_{3\rightarrow4}=T_{4\rightarrow1}=1$
and all remaining matrix elements are 0. An circulator requires an
asymmetric transmission matrix with regard to the dashed line, which
breaks the reciprocity. Conversely, (B) shows a symmetric transmission
matrix measured without control field, representing a reciprocal device.
(C and D) The transmission matrix for the demonstrated circulator
and directional amplifier, respectively. The control power is 7.8
mW for circulator and 6.9 mW for directional amplifier, corresponding
to $C_{\mathrm{cw}}=4.6\textrm{ and }-4.0.$ The values of all transmission
matrix are provided in Supplementary Information. (E) Identical $I$
of circulator, directional amplifier and add-drop filter as a function
of $C_{\mathrm{cw}}$. The lines are the results of theoretical calculations.
}
\end{figure}

The demonstrated non-reciprocal circulator and amplifier based on
the optomechanical interaction in traveling-wave resonator enables
versatile photonic elements, and offers the unique advantages of all-optical
switching, non-reciprocal routing and amplification. Other promising
applications along this direction including the non-reciprocal frequency
conversion, narrowband reflector, as well as synthetic magnetic field
for light by exploiting multiple optical modes in single cavity \cite{OptoMech_Shen,OptoMech3,YanleiZhang}.
With the advances of material and nanofabrication, these devices will
be implemented in photonic integrated circuits \cite{counting}, which
allows stronger optomechanical interaction and smaller device footprint.
Then, the missing block of non-reciprocity could then be tailored
and implemented to meet specific experimental demands. The principle
demonstrated here can also be incorporated into microwave superconducting
devices as well as the acoustic devices in the emerging research field
of quantum phononics \cite{qSAW}.

\emph{Note added: During the preparation of this manuscript, a similar
work by F. Ruesink et al. has been reported on the arXiv} \cite{arXiv},\emph{
where an optical circulator based on microtoroid resonator was demonstrated.
}

\vbox{}

\noindent\textbf{Acknowledgments}\\ The work was supported by the
National Key R\&D Program of China (Grant No.2016YFA0301303, 2016YFA0301700),
the Strategic Priority Research Program (B) of the Chinese Academy
of Sciences (Grant No. XDB01030200), the National Natural Science
Foundation of China (Grant No.61575184, and No. 11722436), the Fundamental
Research Funds for the Central Universities. This work was partially
carried out at the USTC Center for Micro and Nanoscale Research and
Fabrication.

\vbox{}

\noindent\textbf{Author contributions}\\ Z.S., Y.-L.Z. and Y.C.
contribute equally to this work. Z.S., C.-H.D. and C.-L.Z conceived
the experiments, Z.S., Y.C. and C.-H.D. prepared microsphere, built
the experimental setup and carried out measurements. Y.-L.Z., Z.S.
and Y.C. performed the nionumerical simulation and analyzed the data,
X.-B.Z. and F.W.S. provided theoretical support. Z.S., C.-H.D. and
C.-L.Z. wrote the manuscript with input from all co-authors. C.-H.D.,
F.-W.S. and G.-C.G. supervised the project. All authors contributed
extensively to the work presented in this paper\emph{.}

\vbox{}

\noindent\textbf{Additional information}\\ Supplementary information
is available in the online version of the paper. Reprints and permissions
information is available online at. Correspondence and requests for
materials should be addressed to F.-W. S., C.-L. Z or C.-H. D.

\vbox{}

\noindent\textbf{Competing financial interests}\\The authors declare
no competing financial interests.

\end{document}